\documentclass[aps,preprint,superscriptaddress,nofootinbib]{revtex4-1}
\usepackage[T1]{fontenc}
\usepackage[latin9]{inputenc}
\usepackage{verbatim}
\usepackage{amsmath}
\usepackage{amssymb}
\usepackage{graphicx}
\usepackage{esint}
\usepackage[unicode=true,pdfusetitle,
 bookmarks=true,bookmarksnumbered=false,bookmarksopen=false,
 breaklinks=false,pdfborder={0 0 1},backref=false,colorlinks=false]
 {hyperref}

\makeatletter
\@ifundefined{textcolor}{}
{%
 \definecolor{BLACK}{gray}{0}
 \definecolor{WHITE}{gray}{1}
 \definecolor{RED}{rgb}{1,0,0}
 \definecolor{GREEN}{rgb}{0,1,0}
 \definecolor{BLUE}{rgb}{0,0,1}
 \definecolor{CYAN}{cmyk}{1,0,0,0}
 \definecolor{MAGENTA}{cmyk}{0,1,0,0}
 \definecolor{YELLOW}{cmyk}{0,0,1,0}
}

\usepackage{color}

\usepackage{braket}

\@ifundefined{showcaptionsetup}{}{%
 \PassOptionsToPackage{caption=false}{subfig}}
\usepackage{subfig}
\makeatother

\begin{document}

\title{Ionization of hydrogen by neutrino magnetic moment, relativistic
muon, and WIMP}

\author{Jiunn-Wei Chen}
\affiliation{Department of Physics and Center for Theoretical Sciences, National
Taiwan University, Taipei 10617, Taiwan}
\affiliation{National Center for Theoretical Sciences and Leung
Center for Cosmology and Particle Astrophysics, National Taiwan University,
Taipei 10617, Taiwan}

\author{C.-P. Liu}
\affiliation{Department of Physics, National Dong Hwa University, Shoufeng, Hualien
97401, Taiwan}

\author{Chien-Fu Liu}
\affiliation{Department of Physics and Center for Theoretical Sciences, National
Taiwan University, Taipei 10617, Taiwan}

\author{Chih-Liang Wu}
\affiliation{Department of Physics and Center for Theoretical Sciences, National
Taiwan University, Taipei 10617, Taiwan}

\begin{abstract}
We studied the ionization of hydrogen by scattering of neutrino magnetic
moment, relativistic muon, and weakly-interacting massive particle
with a QED-like interaction. Analytic results were obtained and compared
with several approximation schemes often used in atomic physics. As
current searches for neutrino magnetic moment and dark matter have
lowered the detector threshold down to the sub-$\mathrm{keV}$ regime,
we tried to deduce from this simple case study the influence of atomic
structure on the the cross sections and the applicabilities of various
approximations. The general features being found will be useful for
cases where practical detector atoms are considered.
\end{abstract}

\maketitle

\section{Introduction}

The electromagnetic (EM) properties of neutrinos, in particular the magnetic
dipole moments, $\mu_{\nu}$, are of fundamental importance not only in
particle physics but also astrophysics and cosmology (for reviews, see,
e.g., Refs.~\citep{Beringer:1900zz,Broggini:2012df}). In the Standard Model
with massive neutrinos, a non-vanishing $\mu_{\nu}$ arises as a result of
one-loop electroweak radiative correction; for Dirac neutrinos,~\footnote{%
Note that both Dirac and Majorana neutrinos can acquire ``transition''
magnetic dipole moments by similar one-loop radiative corrections in the
Standard Model. In this article we concentrate on the ``static'' ones, which
only Dirac neutrinos can have.} it is given by $\mu_{\nu}=3.20\times10^{-19}%
\,(\frac{m_{\nu}}{\mathrm{eV}})\,\mu_{\mathrm{B}}$, where the Bohr magneton $%
\mu_{\mathrm{B}}=e/(2\, m_{e})$ with $e$ and $m_{e}$ being the magnitude of
charge and mass of electron.~\footnote{%
In this article, we adopt the natural units, $c=\hbar=1$.} From
the current mass upper limit set on the electron neutrino in the tritium $%
\beta$ decay~\citep{Aseev:2011dq}, $m_{\nu_{e}}<2\,\mathrm{eV}$, one can
estimate that $\mu_{\nu_{e}}\lesssim10^{-18}\,\mu_{\mathrm{B}}$ is indeed
very tiny in the Standard Model.

The best direct limits on $\mu_{\nu}$ so far are extracted mostly from
neutrino-electron ($\nu e$) scattering: with the reactor antineutrinos, $%
\mu_{\bar{\nu}_{e}}<2.9\times10^{-11}\,\mu_{\mathrm{B}}$ by the GEMMA
collaboration~\citep{Beda:2013mta} and $\mu_{\bar{\nu}_{e}}<7.4%
\times10^{-11}\,\mu_{\mathrm{B}}$ by the TEXONO collaboration~%
\citep{Wong:2006nx}; with the solar neutrinos, $\mu_{\nu_{\odot}}<5.4%
\times10^{-11}\,\mu_{\mathrm{B}}$ by the Borexino collaboration~%
\citep{Arpesella:2008mt}. Many stronger, but indirect, limits ranging from $%
10^{-11}$ to $10^{-13}$ were inferred from astrophysical or cosmological
constraints, however, they are subject to model dependence and theoretical
uncertainty. Because the current limits, whether direct or indirect, are
orders of magnitude away from the Standard Model prediction, it makes the
search of $\mu_{\nu}$ a powerful probe of new physics.

The cross section of neutrino scattering off a free electron through 
the EM interaction with $\mu _{\nu }$ is \citep{Vogel:1989iv}
\begin{equation}
\left. \frac{d\sigma }{dT}\right\vert _{\mathrm{FE}}=4\,\pi \,\alpha \,\mu
_{\nu }^{2}\,(\frac{1}{T}-\frac{1}{E_{\nu }})\,,  \label{FE}
\end{equation}%
where $\alpha $ is the fine structure constant, $E_{\nu }$ the neutrino
incident energy, and $T$ the neutrino energy deposition. The $1/T$ feature
indicates a way of improving the limit on $\mu _{\nu }$ by lowering the
detector threshold of $T$. Currently the thresholds can be as low as a few $%
\mathrm{keV}$ (e.g., the Germanium semiconductor detectors deployed by both
the GEMMA and TEXONO collaborations), and the next-generation detectors are
geared up to extend down to the sub-$\mathrm{KeV}$ regime~%
\citep{Wong:2011zzd,Yue:2013jja}. While one expects improved limits from
such experimental upgrades, a theoretical issue regarding how the electronic
structure of detectors affects the simple free $\nu e$ scattering formula
naturally arises, as the associated energy scale is comparable to the atomic
scale. Recently there have been discussions about whether atomic structure
can possibly enhance an atomic ionization (AI) cross section~%
\citep{Wong:2010pb,Voloshin:2010vm}, and the robustness of an free electron
approximation in low energy transfer~\citep{Kouzakov:2010tx,Kouzakov:2011vx}%
. With experiments keep pushing down the detector threshold, the need for
more reliable cross section formulae will certainly grow.

Another type of experiments where AI can be relevant is the search for dark
matter (DM), as it shares many similar detection techniques as for $%
\mu_{\nu} $. Most current search focus on the weakly-interacting massive
particles (WIMPs) with masses about $\mathrm{GeV}$ to $\mathrm{TeV}$ scales
-- favoured for astrophysical reasons -- with nuclear recoil in targets being
the main observable. Recently the sub-$\mathrm{GeV}$ DM candidates,
generically classified as light dark matter (LDM), start to get attention~%
\citep{Essig:2011nj}, and the associated AI processes in targets can be used
to constrain the interaction of LDM candidates with electrons and their
masses~\citep{Essig:2012yx}.

Given the importance of understanding the detectors'
response, in particular in low energy regime, our study starts by
considering the simplest atom -- hydrogen. By treating the electrons
as non-relativistic particles and including the one photon exchange together
with the Coulomb interaction, the problem is solved analytically with 
$O(v_{e}^{2})$ and $O(\alpha ^{2})$ errors, where $v_{e}$ 
is the electron velocity. We then compare our result against
various widely-used approximation schemes for the AI through $\mu _{\nu }$
or DM scattering, we try to draw useful information about the
applicabilities of these approximation schemes under various kinematic
conditions. This knowledge serves as a precursor to our currently-ongoing
projects with realistic atomic species.

The article is organized as follows: In Sec.~\ref{sec:formalism}, we lay
down the general formalism for AI cross sections through EM interactions.
The analytic results for the atomic response functions of hydrogen-like
atoms are given explicitly, and approximation schemes including the free 
electron approximation (FEA), equivalent photon approximation (EPA), 
longitudinal photon approximation (LPA), and the one of 
Kouzakov, Studenikin, and Voloshin (KSV)~\citep{Kouzakov:2011vx} 
are introduced. The case of AI by $\mu _{\nu }$ is studied in Sec.~\ref%
{sec:munu_AI}, with particular attention to the issue whether atomic
structure enhances or suppresses the cross sections while scattering occurs
at atomic scales. In Sec.~\ref{sec:muon_AI}, the well-known AI process by
relativistic muon is re-visited. A detailed account of why EPA works for
this case but not for $\mu _{\nu }$ is given. Finally we extend the above
formalism to a QED-like gauge model for the DM interaction with normal
matter, and study the hydrogenic response under various DM kinematics in
Sec.~\ref{sec:DM_AI}. A brief summary is in Sec.~\ref{sec:summary}.

\section{Formalism \label{sec:formalism}}

\begin{figure}[tbp]
\includegraphics[width=0.7%
\textwidth]{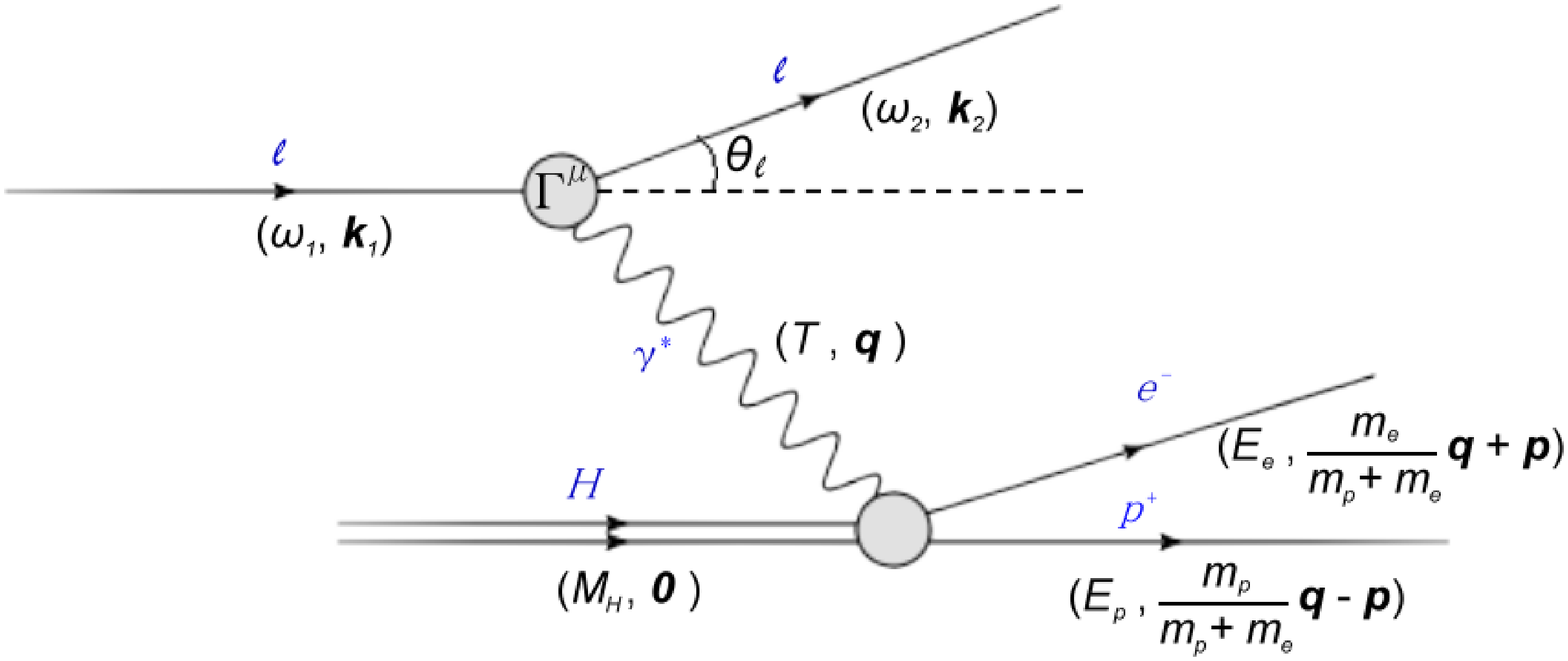}
\caption{The atomic ionization process $l+\mathrm{H}\rightarrow l+\mathrm{H}%
^{+}+e^{-}$ through one photon exchange in the laboratory frame. }
\label{fig:AI}
\end{figure}

Consider the ionization of a hydrogen-like atom $\mathrm{H}$ by a lepton $l$,
\begin{equation}
l+\mathrm{H}\rightarrow l+\mathrm{H}^{+}+e^{-}\,,
\end{equation}%
through one photon exchange, as shown in Fig.~\ref{fig:AI}. We will
treat the electron as a non-relativistic particle and include all its Coulomb
interactions in the initial and final states. 
This problem can be solved analytically. The results will be referred as the
\textquotedblleft full\textquotedblright\ ones -- in comparison to various
approximations to be discussed later on -- and have errors on the order of 
$O(v_{e}^{2},\alpha ^{2})$.

The unpolarized differential cross section in the laboratory frame, i.e.,
the velocity of the incident lepton $\vec{v}_{1}\neq 0$ and the velocity of
the atomic target $\vec{v}_{\mathrm{H}}=0$, is expressed as~\footnote{%
We adopt the normalization $u^{\dagger }\,u=1$ for all Dirac spinors.}

\begin{align}
d\sigma= & \dfrac{1}{|\vec{v}_{1}|}\,\dfrac{(4\,\pi\,\alpha)^{2}}{Q^{4}}\,%
\overline{l}^{\mu\nu}\,\overline{W}_{\mu\nu}\,(2\pi)^{4}\,%
\delta^{4}(k_{1}+p_{\mathrm{H}}-k_{2}-p_{R}-p_{r})\,\dfrac{d^{3}\vec{k}_{2}}{%
(2\,\pi)^{3}}\,\dfrac{d^{3}\vec{p}_{R}}{(2\,\pi)^{3}}\,\dfrac{d^{3}\vec{p}%
_{r}}{(2\,\pi)^{3}}\,,  \label{eq:dsigma-cov}
\end{align}
where the four momenta $k_{1}=(\omega_{1},\vec{k_{1}})$ and $k_{2}=(\omega_{2},%
\vec{k_{2}})$ are of the initial and final leptons, $p_{\mathrm{H}}=(M_{%
\mathrm{H}},\vec{0})$ of the initial atom, $p_{R}=(E_{R},\vec{p}_{R})$ and $%
p_{r}=(E_{r},\vec{p}_{r})$ of the final $\mathrm{H}^{+}+e^{-}$ state in the
center-of-mass and relative coordinates, and $q^{\mu}=k_{1}^{\mu}-k_{2}^{%
\mu}=(T,\vec{q})$ of the virtual photon; respectively; and $%
Q^{2}=q_{\mu}\, q^{\mu}$. The leptonic tensor
\begin{equation}
\overline{l}^{\mu\nu}\equiv\sum_{s_{2}}\overline{\sum_{s_{1}}}%
\braket{k_{2},s_{2}|j_{l}^{\mu}|k_{1},s_{1}}\,\braket{k_{2},s_{2}|j_{l}^{%
\nu}|k_{1},s_{1}}^{*}\,,  \label{eq:l_munu}
\end{equation}
is obtained by a sum of the final spin state $s_{2}$ and an average of the
initial spin state $s_{1}$ of the leptonic electromagnetic (EM) current, $%
j_{l}$, matrix elements; and similarly the atomic tensor
\begin{equation}
\overline{W}^{\mu\nu}\equiv\sum_{m_{j_{f}}}\overline{\sum_{m_{j_{i}}}}%
\braket{f|j_{A}^{\mu}|i}\,\braket{f|j_{A}^{\nu}|i}^{*}\,,  \label{eq:W_muni}
\end{equation}
involves a sum of the final angular momentum state $m_{j_{f}}$ and an
average of the initial angular momentum state $m_{j_{i}}$ of the atomic EM
current, $j_{A}$, matrix elements, where $\ket{i}$ and $\ket{f}$ refer to
atomic initial and final states, respectively.

In this work, we use the relativistic form for $j_{l}^{\mu}$
\begin{equation}
\braket{k_{2},s_{2}|j_{l}^{\mu}|k_{1},s_{1}}=\bar{u}(k_{2},s_{2})%
\,[F_{1}^{(l)}\gamma^{\mu}\,-i\,\frac{F_{2}^{(l)}}{2\, m_{e}}%
\,\sigma^{\mu\nu}\, q_{\nu}]\, u(k_{1},s_{1})\,.  \label{eq:j_leptonic}
\end{equation}
The Dirac and Pauli form factors, $F_{1}^{(l)}$ and $F_{2}^{(l)}$, which
describe the helicity-preserving and helicity-changing EM couplings, are
constant for elementary leptons: $F_{1}^{(l)}$ is the charge $e_{l}$ (in
units of $e$) and $F_{2}^{(l)}$ the anomalous magnetic dipole moment $%
\kappa_{l}$ (in units of $\mu_{\mathrm{B}}$) .

Since we are only interested in the case that the energy deposition
by the incident particle is small enough such that electrons can be treated
as non-relativistic particles, the charge and spatial current densities in
momentum space are 
\begin{eqnarray}
\rho^{(A)}(\vec{q}) &=&-\,e^{i\,\,\vec{q}\cdot \left( \vec{R}+\vec{r}\right)
}\,,  \label{eq:rho_A} \\
\vec{j}^{(A)}(\vec{q}) &=&\frac{-1}{2\,m_{e}}\,e^{i\,\,\vec{q}\cdot \left( \vec{%
R}+\vec{r}\right) }\,(\vec{q}+2\,\vec{p}_{r}+i\,\,\vec{\sigma}_{e}\times
\vec{q}).  \label{eq:j_A}
\end{eqnarray}%
$\rho ^{(A)}$ is leading in the $1/m_{e}$ expansion while 
$\vec{j}^{(A)}$ is subleading. The proton contribution can be
neglected because its contribution to $\vec{j}^{(A)}$ is 
$O(1/m_{p})$ and is smaller than the electron contribution by a
factor of $m_{e}/m_{p}$. Its contribution to $\rho^{(A)}$ 
is smaller than the electron contribution by at least one power of 
$m_{e}/m_{p}$ in the multiple expansion, because the size of the
proton wave function is smaller than that of the electron wave function by
a $m_{e}/m_{p}$ factor.

After performing the spin sum, contraction of the leptonic and atomic
tensors, and implementing the current conservation condition to relate the
longitudinal spatial current to the charge density
\begin{equation}
j_{\parallel}^{(A)}(\vec{q})\equiv\frac{\vec{q}}{q}\cdot\vec{j}^{(A)}(\vec{q}%
)=\frac{T}{q}\,\rho^{(A)}(\vec{q})\,,  \label{eq:j_para-to-rho}
\end{equation}
where $q\equiv|\vec{q}|$, the cross section can be cast into the following
form
\begin{equation}
d\sigma=\frac{\pi}{|\vec{k}_{1}|}\,\dfrac{(4\,\pi\,\alpha)^{2}}{Q^{4}}%
\,\sum_{X=L,T}\left[\left(e_{l}^{2}\, V_{X}^{(F_{1})}+\frac{\kappa_{l}^{2}}{%
(2\, m_{e})^{2}}\, V_{X}^{(F_{2})}\right)\, R_{X}\right]\,\dfrac{d^{3}\vec{k}%
_{2}}{(2\,\pi)^{3}\,2\,\omega_{2}}\,,  \label{eq:dsigma_R}
\end{equation}
through defining the longitudinal and transverse response functions, $R_{L}$
and $R_{T}$, and the corresponding kinematic factors, $V_{L}$ and $V_{T}$.
The kinematic factors, which depend on the energy transfer $T$ and momentum
transfer $q$, are

\begin{align}
V_{L}^{(F_{1})} & =\dfrac{Q^{4}}{q^{4}}[(\omega_{1}+\omega_{2})^{2}-q^{2}],
\label{eq:V_L(F1)} \\
V_{T}^{(F_{1})} & =-\,\left[\dfrac{Q^{2}(Q^{2}+4\omega_{1}\omega_{2})}{2\,
q^{2}}+Q^{2}+2m_{l}^{2}\right]\,,  \label{eq:V_T(F1)}
\end{align}
and

\begin{align}
V_{L}^{(F_{2})} & =\dfrac{-Q^{4}}{q^{4}}\left[(\omega_{1}+\omega_{2})^{2}\,
Q^{2}+4\, m_{l}^{2}\, q^{2}\right],  \label{eq:V_L(F2)} \\
V_{T}^{(F_{2})} & =\dfrac{Q^{2}}{2\, q^{2}}\left[Q^{2}(Q^{2}+4\,\omega_{1}\,%
\omega_{2})-4\, m_{l}^{2}\, q^{2}\right]\,,  \label{eq:V_T(F2)}
\end{align}
for couplings with the $F_{1}^{(l)}$ and $F_{2}^{(l)}$ form factors,
respectively.~\footnote{%
As $Q^{2}=T^{2}-q^{2}$ and $\omega_{2}=\omega_{1}-T$, the independent
variables in these expressions are thus taken by $(T,q)$.} The response
functions, which are also functions of $(T,q)$ but independent of the form
of leptonic coupling, are
\begin{eqnarray}
R_{L} & \equiv & \sum_{m_{j_{f}}}\overline{\sum_{m_{j_{i}}}}\int\,\frac{d^{3}%
\vec{p}_{r}}{(2\,\pi)^{3}}\,|\langle f|\rho^{(A)}(\vec{q})|i\rangle|^{2}\delta%
\left(T-B-\dfrac{q^{2}}{2M}-\dfrac{p_{r}^{2}}{2\,\mu_{red}}%
\right)\,,  \label{eq:R_L_def} \\
R_{T} & \equiv & \sum_{m_{j_{f}}}\overline{\sum_{m_{j_{i}}}}\int\,\frac{d^{3}%
\vec{p}_{r}}{(2\,\pi)^{3}}\,|\langle f|j_{\perp}^{(A)}(\vec{q})|i\rangle|^{2}%
\delta\left(T-B-\dfrac{q^{2}}{2M}-\dfrac{p_{r}^{2}}{2\,\mu_{red}%
}\right)\,,  \label{eq:R_T_def}
\end{eqnarray}
where $B$ is the binding energy of the $\mathrm{H}$ atom, 
$M=m_{e}+m{}_{p}\approx m_{p}$, and $\mu_{red}=m_e\,m_p/(m_e+m_p)\approx m_e$. 
Note that the center-of-mass degrees
of freedom in the final state have been integrated out by the momentum
conservation, which yield $\vec{p}_{R}=\vec{q}$; and the resulting energy
conservation delta function properly takes care the nuclear recoil effect.

Consider now the ionization of a hydrogen-like atom from its ground state,
i.e., the $1s$ orbit, the relevant atomic spatial wave functions for the
initial ($i$) and final ($f$) states are
\begin{eqnarray}
\braket{\vec{r}|i} &=&\braket{\vec{r}|(nlm_{l}=100)}=\frac{1}{\sqrt{\pi }}%
\,Z^{3/2}\,e^{-Z\,\bar{r}}\,,  \label{eq:psi_i} \\
\braket{f|\vec{r}} &=&^{(-)}\braket{\vec{p}_{r}|\vec{r}}=e^{\frac{\pi \,Z}{%
2\,\bar{p}_{r}}}\Gamma \left( 1-\dfrac{i\,Z}{\bar{p}_{r}}\right) \,e^{-i\,%
\vec{p}_{r}\cdot \vec{r}}{_{1}F_{1}}\left( \dfrac{i\,Z}{\bar{p}_{r}}%
,1,i(p_{r}\,r+\vec{p}_{r}\cdot \vec{r})\right) \,,  \label{eq:psi_f}
\end{eqnarray}%
in atomic units (so the barred quantities are $\bar{r}=r\,m_{e}\,\alpha $, $%
\bar{p}_{r}=p/(m_{e}\,\alpha )$, etc.), where $\Gamma (z)$ and ${_{1}F_{1}}%
(a,b,z)$ are the Gamma and confluent hypergeometric functions, respectively.
The evaluations of $R_{L}$ and $R_{T}$ can be done analytically by the
Nordsieck integration~%
\citep{Nordsieck:1953aa,Holt:1969ar,Belkic:1981dz,Gravielle:1991mi} and
yield:
\begin{align}
R_{L}& =\frac{2^{8}\,Z^{6}\,\bar{q}^{2}\,(3\,\bar{q}^{2}+\bar{p}%
_{r}^{2}+Z^{2})\,\exp \left[ -\frac{2\,Z}{\bar{p}_{r}}\tan ^{-1}\left( \frac{%
2\,Z\,\bar{p}_{r}}{\bar{q}^{2}-\bar{p}_{r}^{2}+Z^{2}}\right) \right] }{3\,((%
\bar{q}+\bar{p}_{r})^{2}+Z^{2})^{3}\,((\bar{q}-\bar{p}_{r})^{2}+Z^{2})^{3}%
\,(1-e^{-2\,\pi \,Z/\bar{p}_{r}})}  \label{eq:R_L_1s} \\
R_{T}& =\frac{2^{7}\,\alpha ^{2}\,Z^{6}\,(\bar{p}_{r}^{2}+Z^{2})\,\exp \left[
-\frac{2\,Z}{\bar{p}_{r}}\tan ^{-1}\left( \frac{2\,Z\,\bar{p}_{r}}{\bar{q}%
^{2}-\bar{p}_{r}^{2}+Z^{2}}\right) \right] }{3\,((\bar{q}+\bar{p}%
_{r})^{2}+Z^{2})^{2}\,((\bar{q}-\bar{p}_{r})^{2}+Z^{2})^{2}\,(1-e^{-2\,\pi
\,Z/\bar{p}_{r}})}+\frac{1}{2}\,\mu _{e}^{2}\,\alpha ^{2}\,\bar{q}%
^{2}\,R_{L}\,.  \label{eq:R_T_1s}
\end{align}%
The first term in $R_{T}$ is the contribution from the convection current,
and the second one from the spin current. The overall $\alpha ^{2}$ factor
appearing in both terms reflects the $1/m_{e}^{2}$ order in the
non-relativistic expansion. In comparison, $R_{L}$ is $O(\alpha ^{0})$.

The single differential cross section with respect to the energy transfer
can then be computed by integration over the lepton scattering angle $\theta$
\begin{align}
\dfrac{d\sigma}{dT} & =\int\, d\cos\theta\,\dfrac{2\,\pi\,\alpha^{2}}{Q^{4}}%
\dfrac{k_{2}}{k_{1}}\left(V_{L}\, R_{L}+V_{T}\, R_{T}\right)
\label{eq:dsigma/dT} \\
V_{L,T} & =e_{l}^{2}\, V_{L,T}^{(F_{1})}+\frac{\kappa_{l}^{2}}{(2\,
m_{e})^{2}}\, V_{L,T}^{(F_{2})}  \notag
\end{align}
with a constrained range of $\cos\theta$:
\begin{align}
\min\left\{ 1,\max\left[-1,\dfrac{k_{1}^{2}+k_{2}^{2}-2\, M_{H}\,(T-B)}{2\,
k_{1}\, k_{2}}\right]\right\} \leq\cos\theta\leq1.  \label{cos range}
\end{align}
For latter discussion, we note that for a fixed energy transfer, the square
of four momentum transfer:

\begin{equation}
Q^{2}=2\, m_{l}^{2}-2\,\omega_{1}\,(\omega_{1}-T)+2\,\sqrt{%
\omega_{1}^{2}-m_{l}^{2}}\,\sqrt{(\omega_{1}-T)^{2}-m_{l}^{2}}\,\cos\theta\,,
\label{eq:Qsquared}
\end{equation}
only depends on $\cos\theta$, therefore, the integration over $\cos\theta$
is equivalent of integrating over $Q^{2}$ (or $q^{2}$).

While it is straightforward to obtain complete and analytic results for
ionizations of the hydrogen atom to the order outlined above,~%
\footnote{%
Ionizations of hydrogen-like atoms in metastable states can be performed
similarly, however, the analytic results get more and more tedious as the
principle quantum number $n$ grows.} we shall discuss several approximation
schemes often employed in atomic calculations, and compare them with the
full calculations for this case study in the following sections.

\subsection{Free Electron Approximation (FEA)}

The FEA is expected be a good approximation if the photon wavelength is
much smaller than the size of the atom (or the typical distance between
electrons in a multi-electron system) such that the atomic effect is no
longer important. Thus, a necessary (but not sufficient) condition for this
approximation to be valid is that the scattering energy needs to be high
(compared with the typical scale of the problem).

In this approximation, the electron before and after ionization is treated 
as a free particle. The free electron cross section of Eq.(\ref{FE}) is
multiplied by the step function $\theta (T-B)$ to incorporate the 
binding effect: 
\begin{equation}
\left. \dfrac{d\sigma }{dT}\right\vert _{FEA}=\theta (T-B)\left. \dfrac{%
d\sigma }{dT}\right\vert _{FE}\,.
\end{equation}%
Energy and momentum conservation fixes $Q^{2}=-2\,m_{e}\,T$ in this two-body phase space.

\subsection{Equivalent Photon Approximation (EPA)}

The equivalent photon approximation~\citep{Weizsacker:1934sx,Williams:1934ad} 
treats the virtual photon as a real (and thus transversely polarized) photon. 
It could be a good approximation for low energy processes where the photon is soft such
that $Q^{2}\approx 0$ because $q^{\mu }\approx 0$ for
every component of $\mu$. At high energies, besides soft
photon emissions, when the initial and final state electrons are highly
relativistic and almost collinear, the emitted ``collinear'' photon also has 
$Q^{2}\approx 0$.
While the soft photon emission is likely to dominate
the phase space of low energy scattering, whether the soft and collinear
photon emission will dominate the high energy scattering depends on the
transition matrix elements.

The total cross section $\sigma _{\gamma }$ for the photoionization process $%
\gamma +\mathrm{H}\rightarrow \mathrm{H}^{+}+e^{-}$ is

\begin{equation}
\sigma _{\gamma }(T)=\dfrac{2\,\pi ^{2}\,\alpha }{T}\,R_{T}^{0}\,,
\end{equation}%
where the photon energy $E_{\gamma }=T$ and the superscript
\textquotedblleft $0$\textquotedblright\ denotes that the photon is
\textquotedblleft on-shell\textquotedblright , i.e., $T^{2}=q^{2}$. Then EPA
relates $\sigma _{\gamma }$ to a corresponding lepto-ionization process
(involving a virtual photon) by the following two steps: (i) ignoring the
longitudinal response function $R_{L}$ and (ii) substituting the off-shell
response function $R_{T}$ by the on-shell $R_{T}^{0}$ extracted from the
photo-ionization process, i.e.,
\begin{align}
\left. \dfrac{d\sigma }{dT}\right\vert _{\mathrm{EPA}}& =\int \,d\cos \theta
\,\dfrac{2\,\pi \,\alpha ^{2}}{Q^{4}}\dfrac{k_{2}}{k_{1}}\left[ V_{T}\left(
\dfrac{T}{2\,\pi ^{2}\,\alpha }\,\sigma _{\gamma }(T)\right) \right] \,,
\notag \\
& \equiv \frac{1}{T}\,N(T)\,\sigma _{\gamma }(T)\,,  \label{eq:dsigma/dT-EPA}
\end{align}%
with the energy spectrum of equivalent photon $N(T)$ defined by
\begin{equation}
N(T)=\frac{\alpha }{\pi }\,\frac{k_{2}}{k_{1}}\,T^{2}\,\int \,d\cos \theta \,%
\frac{V_{T}}{Q^{4}}\,,
\end{equation}%
where the integration range of $\cos \theta $ is the same as Eq.~(\ref{cos
range}). Because it directly feeds the photo-ionization cross sections
(experimental accessible) to the corresponding lepto-ionization cross
sections, a lot of theoretical work and uncertainties can be saved when it
works properly.

At this point, we should make an important remark as regards the
approximation scheme adopted in Ref.~\citep{Wong:2010pb}: Even though it is
in the spirit of the EPA, however, it makes a stronger assumption that the
integration leading to energy spectrum of equivalent photon is also
dominated by the $Q^{2}\approx 0$ region (or staying constant), i.e.,
\begin{equation}
\left. N(T) \right \vert_\mathrm{EPA^{\ast}} \approx 
\frac{\alpha }{\pi }\,\frac{k_{2}}{k_{1}}\,T^{2}\,\int \,d\cos
\theta \,\left. \frac{V_{T}}{Q^{4}}\right\vert _{Q^{2}\approx 0}\,.
\end{equation}%
To distinguish this stronger version of the EPA from the conventional one,
we shall denote it as the EPA$^{\ast }$ scheme.

\subsection{Longitudinal Photon Approximation (LPA)}

The longitudinal photon contribution is leading order in the 
$1/m_{e}$ expansion while the transverse photon contribution is
subleading. Thus, it might be a good approximation for non-relativistic
systems:
\begin{equation}
\left. \dfrac{d\sigma }{dT}\right\vert _{\mathrm{LPA}}=\int \,d\cos \theta \,%
\dfrac{2\,\pi \,\alpha ^{2}}{Q^{4}}\,\dfrac{k_{2}}{k_{1}}\,V_{L}\,R_{L}\,.
\end{equation}%
The difference of this approximation to the full calculation is a measure of
how importantly the transverse current contributes to the process.

\subsection{Approximation Scheme of Kouzakov, Studenikin, and Voloshin (KSV)}

The KSV scheme includes the longitudinal photon contribution which
is leading order in the $1/m_{e}$ expansion and approximates the
subleading transverse photon contribution by a relation only strictly suitable for
the electric dipole ($E_{1}$) transition in the long wavelength limit, i.e., $q\rightarrow 0$: 
\begin{equation}
R_{T}=2\,\frac{T^{2}}{q^{2}}\,R_{L}\,.\,\footnote{Note that the factor of 2 difference from Eq.~(12) of 
Ref.~\citep{Kouzakov:2011vx} is due to the different definitions for the transverse response function.}  
\label{eq:Siegert_Thm}
\end{equation}%

This relation can be derived from the Siegert theorem~\citep{Siegert:1937yt} 
for $E_{1}$, which is based on current conservation. It can also be
explicitly checked by taking the same limit to Eqs.~(\ref{eq:R_L_1s},\ref%
{eq:R_T_1s}). In general $R_{T}$ is not dominated by $E_{1}$ in the processes
that we are considering, which requires $q\,r_{A}\ll 1$ where $\,r_{A}$ is
the size of the atom. But Eq.~\ref{eq:Siegert_Thm} can still be a good approximation 
to the cross section calculations as long as $R_{T}$ remains subleading to $R_{L}$.

In Refs.~\citep{Kouzakov:2010tx,Kouzakov:2011vx}, the authors adopted the
above relation so the cross section were calculated without need to evaluate
the transverse response function which is harder to compute:
\begin{equation}
\left. \dfrac{d\sigma }{dT}\right\vert _{\mathrm{KSV}}=\int \,d\cos \theta \,%
\dfrac{2\,\pi \,\alpha ^{2}}{Q^{4}}\dfrac{k_{2}}{k_{1}}\left( V_{L}+2\,\frac{%
T^{2}}{q^{2}}\,V_{T}\right) R_{L}\,.
\end{equation}

\section{Ionization by Neutrino Magnetic Moment \label{sec:munu_AI}}

\begin{figure}
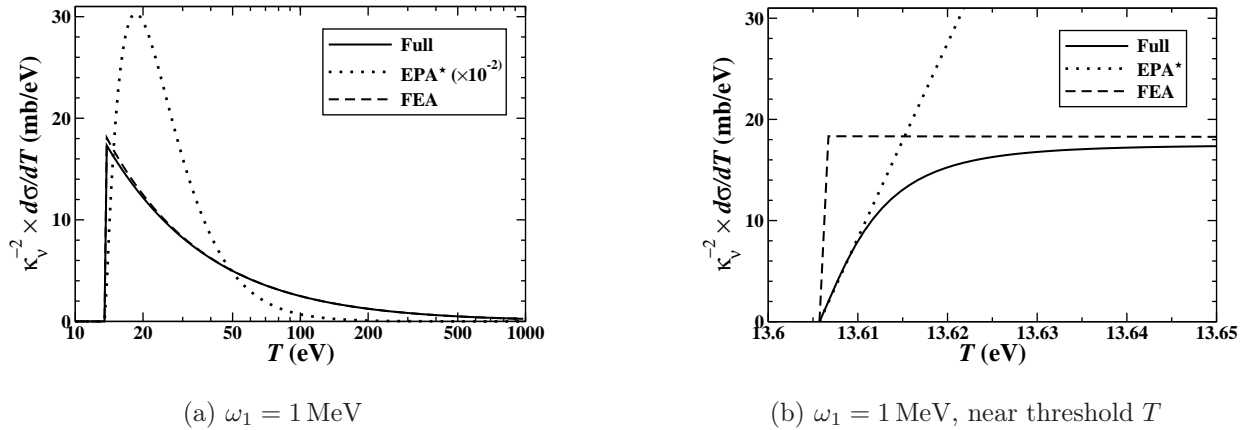

\subfloat[$\omega_1=1\,\mathrm{MeV}$ \label{fig:dsigma/dT_nu_full}]
{\includegraphics[scale=1]{nu-H_AI_w1_1MeV}}
\hfill{}
\subfloat[$\omega_1=1\,\mathrm{MeV}$, near threshold $T$ \label{fig:dsigma/dT_nu_th}]
{\includegraphics[scale=1]{nu-H_AI_w1_1MeV_th}}

\caption{Differential cross sections $\frac{d\sigma}{dT}$ for $\bar{\nu}+\mathrm{H}\rightarrow\bar{v}+p+e^{-}$
via the EM interaction with the neutrino magnetic moment $\mu_{\nu}=\kappa_{\nu}\,\mu_{\mathrm{B}}$.
The incident neutrino has energy $\omega_{1}=1\,\mathrm{MeV}$ with
its mass $m_{\nu}$ taken to be zero. The results of the approximation
schemes KSV and LPA (both not shown) are in excellent agreement with the
full calculation. \label{fig:dsigma/dT_nu}}
\end{figure}

In case the incident lepton is a neutrino ($\nu$) or antineutrino
($\bar{\nu}$), as $e_{\nu}=0$, the EM breakup process is thus sensitive
to the neutrino magnetic moment $\mu_{\nu}=\kappa_{\nu}\,\mu_{\mathrm{B}}$
(which is purely anomalous). The energy spectrum for reactor antineutrinos
typically peaks around few tens of $\mathrm{keV}$ to $\mathrm{MeV}$
(see, e.g., Ref.~\citep{Wong:2006nx}); setting $\omega_{1}=1\,\mathrm{MeV}$,
a plot of the single differential cross section with energy loss up
to $1\,\mathrm{keV}$ is given in Fig.~\ref{fig:dsigma/dT_nu}. Not
shown in these figures are the results of the approximation schemes
KSV and LPA. They both agree with the full calculation to good extents:
within $10^{-5}$ for the former and $10^{-3}$ for the latter in
the entire range. In other words, this atomic bound-to-free transition
is dominated by the atomic charge operator, while the transverse current
operator is negligible, which implies the inadequacy of the EPA scheme.

The dominance of the charge operator over the transverse current can
be roughly understood by a comparison of their corresponding kinematic
factors $V_{L}^{(F_{2})}$ and $V_{T}^{(F_{2})}$. For neutrino scattering
\begin{equation}
Q^{2}|_{m_{\nu=0}}\approx-2\,\omega_{1}^{2}(1-x)\,,
\end{equation}
where $x \equiv \cos\theta $, they are
\begin{align}
\frac{V_{L}^{(F_{2})}}{Q^{4}}=\frac{2\,(1-x)}{(1-x+\frac{T^{2}}{2\,\omega_{1}^{2}})^{2}}\,, & \qquad\frac{V_{T}^{(F_{2})}}{Q^{4}}=\frac{(1+x)}{2\,(1-x+\frac{T^{2}}{2\,\omega_{1}^{2}})}\,.
\end{align}
As $T^{2}/\omega_{1}^{2}\ll1$ in our consideration, both functions
peak near $x=1$, and they have similar maximum values: $V_{L}^{(F_{2})}/Q^{4}|_{\mathrm{max}}=\omega_{1}^{2}/T^{2}$
and $V_{T}^{(F_{2})}/Q^{4}|_{\mathrm{max}}=2\,\omega_{1}^{2}/T^{2}$, and widths.
Since the transverse response function does not get enhancement from
the kinematic factor $V_{T}^{(F_{2})}$ over $V_{L}^{(F_{2})}$, its
contribution to the cross section is suppressed by the usual non-relativistic
order $\alpha^{2}$.

The good agreements with the FEA scheme (Fig.~\ref{fig:dsigma/dT_nu_full})
is not really a surprise: the energetic neutrino emits a virtual photon
with wavelength smaller than the atomic size so that the binding effect
does not manifest in a short distance. The only exception is near
the ionization threshold (Fig.~\ref{fig:dsigma/dT_nu_th}) where
the virtual photon wavelength is larger than the atomic size, and
the binding effect suppresses the cross section in comparison to FEA.

\begin{figure}
\subfloat[$\omega_{1}=1\,\mathrm{MeV}$, $T=20\,\mathrm{eV}$ 
\label{fig:dsigma/dTdx_nu_w1_l}]
{\includegraphics[scale=1]{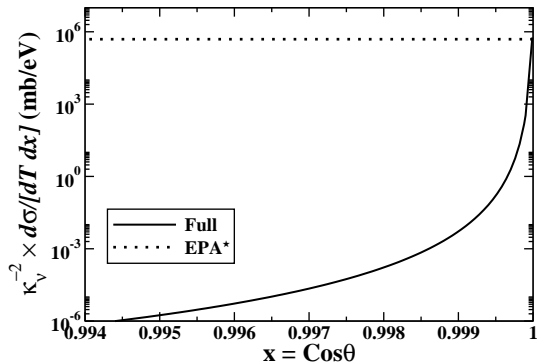}}
\hfill{}
\subfloat[$\omega_{1}=1\,\mathrm{keV}$, $T=20\,\mathrm{eV}$ \label{fig:dsigma/dTdx_nu_w1_s}]
{\includegraphics[scale=1]{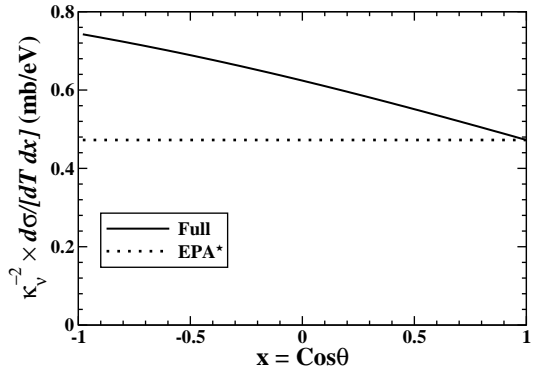}}

\caption{Double differential cross sections $\frac{d\sigma}{dT\, d\cos\theta}$
for $\bar{\nu}+\mathrm{H}\rightarrow\bar{v}+p+e^{-}$ via the EM interaction
with the neutrino magnetic moment $\mu_{\nu}$. The energy transfer
is fixed at $20\,\mathrm{eV}$. EPA$^{*}$ hugely overestimates at high $\omega_1$, but works reasonably at low $\omega_1$. \label{fig:dsigma/dTdx_nu}}
\end{figure}

Also shown in these figures is the result of the EPA$^{*}$ scheme.
Note that the curve in Fig.~\ref{fig:dsigma/dT_nu_full} has to be
scaled down by a factor of $100$ in order to be cast on the same
plot as other calculations; in other words, the EPA$^{*}$ hugely
overestimates the cross section by several orders of magnitude. There
is only a tiny region near the ionization threshold (see Fig.~\ref{fig:dsigma/dT_nu_th})
where the EPA$^{*}$ does work; that is where the virtual photon approaches
the real photon limit ($T=q$). The origin of such an overestimate
can be clearly seen in Fig.~\ref{fig:dsigma/dTdx_nu_w1_l}, where
the double differential cross section $d\sigma/(dT\, dx)$ 
is plotted as a function of $x$ with a fixed
energy loss $T=20\,\mathrm{eV}$. Due to the kinematic constraint,
the maximum scattering angle $\theta_{max}=6.28^{\circ}$. However,
even within this small range of peripheral scattering angle, the differential
cross section decreases dramatically by $12$ orders of magnitude
from the forward angle as a combined result of the kinematic factors
$V_{L,T}$ and the response functions $R_{L,T}$. Therefore, the flatness
of $d\sigma/(dT\, dx)$ required by the EPA$^{*}$ is severely violated
and results in this overestimation.

\begin{figure}
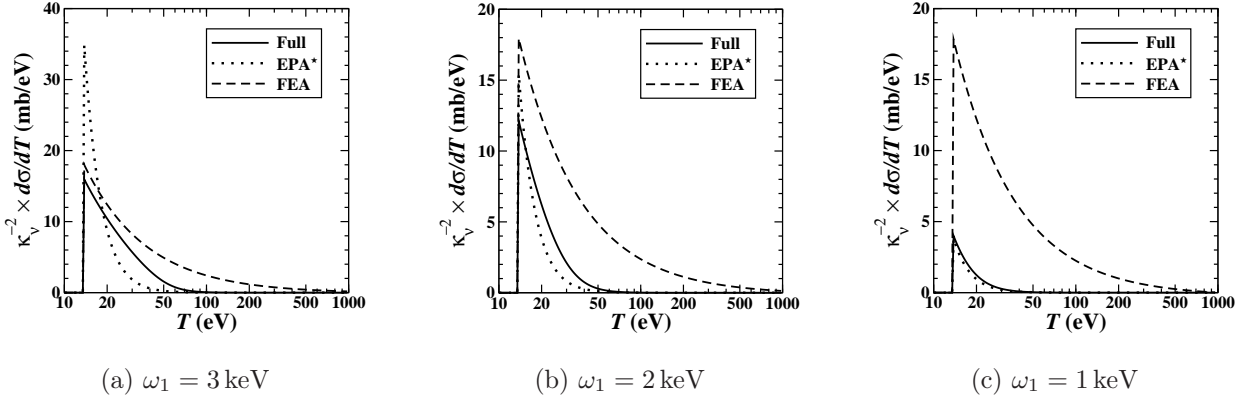

\subfloat[$\omega_1=3\,\mathrm{keV}$]{\includegraphics[scale=0.9]{nu-H_AI_w1_3keV}}
\hfill{}
\subfloat[$\omega_1=2\,\mathrm{keV}$]{\includegraphics[scale=0.9]{nu-H_AI_w1_2keV}}
\hfill{}
\subfloat[$\omega_1=1\,\mathrm{keV}$]{\includegraphics[scale=0.9]{nu-H_AI_w1_1keV}}

\caption{Differential cross sections $\frac{d\sigma}{dT}$ for $\bar{\nu}+\mathrm{H}\rightarrow\bar{v}+p+e^{-}$
via the EM interaction with the neutrino magnetic moment $\mu_{\nu}$
at few $\mathrm{keV}$ incident energies. The EPA$^{*}$ calculations
gradually converge to the full ones. \label{fig:dsigma/dT_nu_low}}
\end{figure}

On the other hand, one does see the EPA$^{*}$ start to work when
the incident neutrino energy $\omega_{1}$ drops below the binding
momentum of the hydrogen-like atom $\sim Z\, m_{e}\,\alpha$. For
hydrogen, the scale is about $3.73\,\mathrm{keV}$, and Fig.~\ref{fig:dsigma/dT_nu_low}
shows that varying $\omega_{1}$ from $3\,\mathrm{keV}$, $2\,\mathrm{keV}$,
to $1\,\mathrm{keV}$, the EPA$^{*}$ result becomes reasonably good.
As evidenced from Fig.~\ref{fig:dsigma/dTdx_nu_w1_s}, the double
differential cross section for $\omega_{1}=1\,\mathrm{keV}$ and $T=20\,\mathrm{eV}$,
even though not looking completely flat, does vary only modestly with
increasing $\theta$, and the agreement is getting better when $\omega_{1}$
is further decreased. In the meanwhile, the FEA is no longer a good
approximation since the de Broglie wavelength of the incident neutrino
is on the order of atomic size so the binding effect is not negligible.

\section{Ionization by Muon \label{sec:muon_AI}}

\begin{figure}
\subfloat[$\omega_1=1\,\mathrm{GeV}$]{\includegraphics[scale=1]{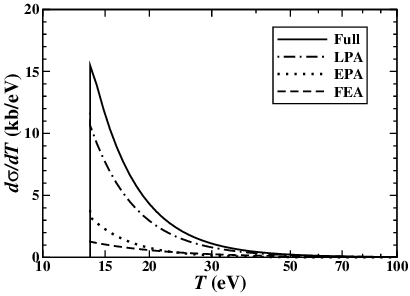}}
\hfill
\subfloat[$\omega_1=10\,\mathrm{GeV}$]{\includegraphics[scale=1]{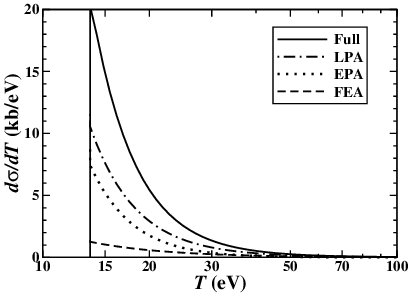}}\\
\vfill \vspace{0.5cm}
\subfloat[$\omega_1=100\,\mathrm{GeV}$]{\includegraphics[scale=1]{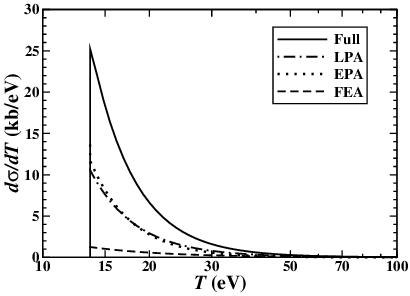}}
\hfill
\subfloat[$\omega_1=1000\,\mathrm{GeV}$]{\includegraphics[scale=1]{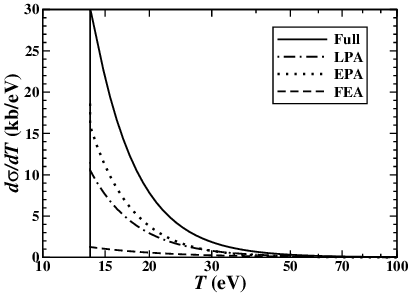}}

\caption{Differential cross sections $\frac{d\sigma}{dT}$ for $\mu^{-}+\mathrm{H}\rightarrow\mu^{-}+p+e^{-}$
via the EM interaction. The results of the approximation schemes KSV
(not shown) are in excellent agreement with the full calculation.\label{fig:dsigma/dT_muon}}
\end{figure}

Replacing the incident lepton from a neutrino to a muon $(\mu^{-})$,
as $e_{\mu^{-}}=-1$, the EM breakup process is instead dominated
by the $F_{1}$ coupling, while the $F_{2}$ coupling can be ignored
for the smallness of muon $g-2\approx0.001$~\citep{Beringer:1900zz}.
Consider relativistic muons with $10^{0,1,2,3}\,\mathrm{GeV}$ energies,
the differential cross sections are plotted in Fig.~\ref{fig:dsigma/dT_muon}.
(Because the muon is relativistic while the electron non-relativistic,
the final state interaction between the muon and electron can be ignored,
see, e.g., Ref.~\citep{Landau:NQM}.) The noticeable differences
in comparison to what have been drawn in the previous neutrino case
are: (1) The differential cross section falls off more quickly as
$T$ increases, i.e., the recoil electrons tend to have relatively
smaller energies. (2) The FEA results are insensitive to $\omega_{1}$
and largely underestimates in all cases. (3) There are substantial
contributions from the transverse current, despite its $ $built-in
$O(Z^{2}\,\alpha^{2})$ suppression due to the non-relativistic kinematics
of atomic electrons. In fact, when $\omega_{1}$ becomes big enough,
the interaction with the atomic transverse current dominates over
the one with the charge and longitudinal current, as indicated by
the competition between the EPA and the LPA curves in Fig.~\ref{fig:dsigma/dT_muon},
and one expects the larger $\omega_{1}$ increases, the better the
EPA works.

The reason for such differences is primarily due to the associated
kinematic factors. At the $Q^{2}\rightarrow0$ limit, they behave
like $V_{L}^{(F_{1})}\propto Q^{4}$ and $V_{T}^{(F_{1})}\propto Q^{0}$
for muon ($m_{\mu}\neq0$), and $V_{L}^{(F_{2})}\propto Q^{6}$
and $V_{T}^{(F2)}\propto Q^{4}$ for neutrino ($m_{\nu}\approx0$)
ionization, respectively. As the differential cross section $d\sigma/dT$
involves an $1/Q^{4}$ weighted integration over $Q^{2}$, only the
transverse part in muon ionization receives a strong weight at peripheral
scattering angles (where $Q^{2}\approx0$).This explains the importance
of the transverse current in relativistic muon ionization and its
insignificance in neutrino ionization. Also, with the allowed scattering
angles become closer to the exact forward direction as the muon incident
energy increases (with $T$ fixed), the kinematics becomes real-photon-like
and eventually the huge enhancement by the $1/Q^{4}$ weight is able
to overcome the non-relativistic suppression in the transverse response
function. The failure of the FEA in relativistic muon ionization can
also be understood in a similar way: In the FEA scheme, the differential
cross section $d\sigma/dT$ is determined from a specific kinematics
$Q_{\mathrm{FEA}}^{2}=-2\, m_{e}\, T$ by energy-momentum conservation;
on the other hand, the full calculation with two-body kinematics involves
an integration over allowed $|Q^{2}|$ ranging from $\approx0$ to
some maximum value determined by the maximum scattering angle. Because
the $1/Q^{4}$ factor that enhances the contributions from the $Q^{2}\approx0$
region, the $Q_{\mathrm{FEA}}^{2}=-2\, m_{e}\, T$ ceases to be a
good representative point.

A semi-quantitative understanding could be obtained by the following
approximate forms of $V_{L}^{(F_{1})}$ and $V_{T}^{(F_{1})}$. For
a relativistic muon
\begin{equation}
Q^{2}|_{\omega_{1}\gg m_{\mu}}\approx-2\,\omega_{1}^{2}\,(1-x)-m_{\mu}^{2}\,\frac{T^{2}}{\omega_{1}^{2}}\,,
\end{equation}
they are
\begin{align}
\frac{V_{L}^{(F_{1})}}{Q^{4}} & =\frac{1}{2\,\omega_{1}^{2}}\,\frac{(1+x)}{(1-x+\frac{T^{2}}{2\,\omega_{1}^{2}})^{2}}\,,\nonumber \\
\frac{V_{T}^{(F_{1})}}{Q^{4}} & =\frac{1}{4\,\omega_{1}^{2}}\,\frac{(1-x)(3-x)+(3+x)\,\frac{m_{\mu}^{2}}{\omega_{1}^{2}}\,\frac{T^{2}}{2\,\omega_{1}^{2}}}{(1-x+\frac{T^{2}}{2\,\omega_{1}^{2}})(1-x+\frac{m_{\mu}^{2}}{\omega_{1}^{2}}\,\frac{T^{2}}{2\,\omega_{1}^{2}})^{2}}\,.
\end{align}
One sees that unlike the previous case for which $V_{T}^{(F_{2})}$
and $V_{L}^{(F_{2})}$ are comparable in most range of $x$, $V_{T}^{(F_{1})}/Q^{4}$
is comparable to $V_{L}^{(F_{1})}/Q^{4}$ only for $1-x\gtrsim\frac{T^{2}}{2\,\omega_{1}^{2}}$.
As the scattering angle further decreases, $V_{T}^{(F_{1})}/Q^{4}$
starts to dominate over $V_{L}^{(F_{1})}/Q^{4}$, and when $1-x\lesssim\frac{m_{\mu}^{2}}{\omega_{1}^{2}}\,\frac{T^{2}}{2\,\omega_{1}^{2}}$,
it overwhelms by a factor $\frac{\omega_{1}^{2}}{m_{\mu}^{2}}\gg1$.
Also because of this huge weight on extremely small angles, the FEA
scheme with $|Q_{\mathrm{FEA}}^{2}|=2\, m_{e}\, T$ overestimates
the averaged $\overline{|Q^{2}|}$ for the realistic situation and
leads to an underestimation.

\begin{figure}

\includegraphics[scale=1.2]{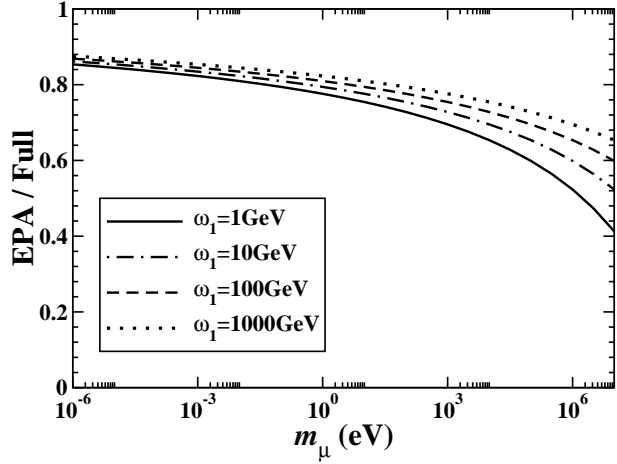}

\caption{The EPA scheme as an approximation for the relativistic muon ionization
with an adjustable $m_{\mu}$, with $T=15\,\mathrm{eV}$. \label{fig:EPA_mass_dep} }
\end{figure}

Though one sees that the EPA serves as a better approximation than
the FEA in the relativistic muon ionization, however, as shown in
Fig.~\ref{fig:dsigma/dT_muon}, even at $\omega_{1}=1000\,\mathrm{GeV}$,
it can still not be taken as a good approximation to the full result.
The main reason is its non-zero mass which limits the lowest |$Q^{2}|$
to be reached
\begin{equation}
|Q^{2}|_{\min}\approx\frac{T^{2}}{\omega_{1}^{2}}\, m_{\mu}^{2}|_{\omega_{1}\gg m_{\mu}\gg T}\,.
\end{equation}
If $m_{\mu}$ is adjusted to smaller values $\lesssim1\,\mathrm{eV}$,
then indeed the EPA accounts for $\gtrsim80\%$ of the cross section
for $\omega_{1}$ on the orders of $\mathrm{GeV}-\mathrm{TeV}$, as
shown in Fig.~\ref{fig:EPA_mass_dep}. In other words, in case one
seeks a better description of relativistic muon ionization or other
processes alike beyond the EPA, the contribution from charge and longitudinal
current should be included.

\section{Ionization by WIMP \label{sec:DM_AI}}

\begin{figure}
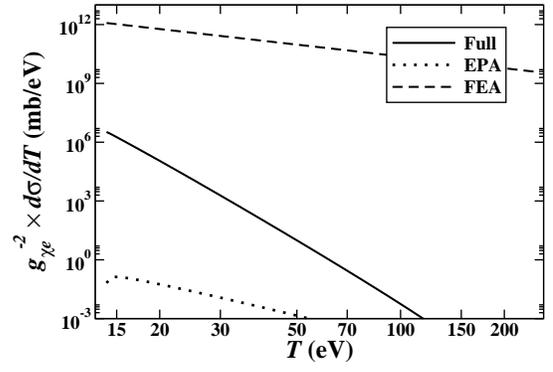
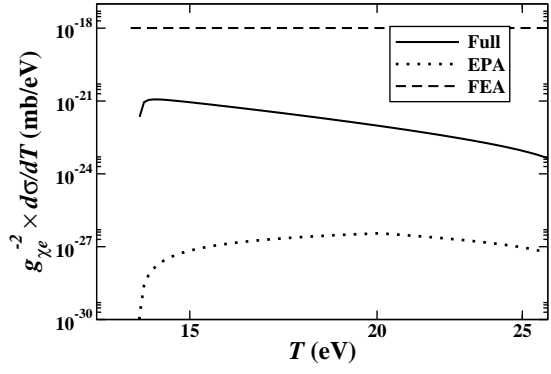
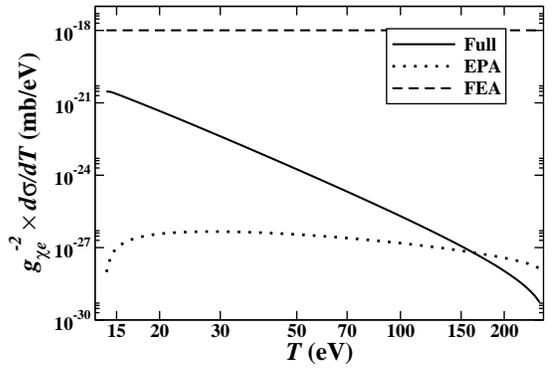
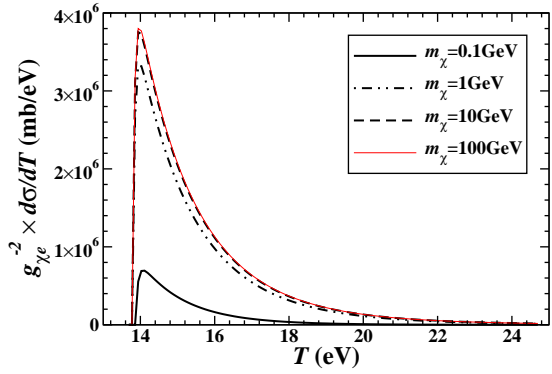
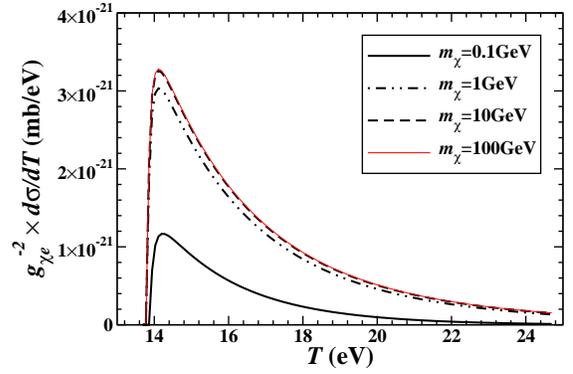

\subfloat[$m_{\chi}=0.1\,\mathrm{GeV}$, $m_b=0$]{\includegraphics[scale=1]{DM-H_AI_F1_QED_100MeV}}
\hfill
\subfloat[$m_{\chi}=1\,\mathrm{GeV}$, $m_b=0$]{\includegraphics[scale=1]{DM-H_AI_F1_QED_1GeV}}
\\ \vspace{0.5cm}
\subfloat[$m_{\chi}=0.1\,\mathrm{GeV}$, $m_b=125\,\mathrm{GeV}$]
{\includegraphics[scale=1]{DM-H_AI_F1_4F_100MeV}}
\hfill
\subfloat[$m_{\chi}=1\,\mathrm{GeV}$, $m_b=125\,\mathrm{GeV}$]{\includegraphics[scale=1]{DM-H_AI_F1_4F_1GeV}}
\\ \vspace{0.5cm}
\subfloat[$m_b=0$, near threshold]
{\includegraphics[scale=1]{DM-H_AI_F1_QED_th}}
\hfill
\subfloat[$m_b=125\,\mathrm{GeV}$, near threshold]
{\includegraphics[scale=1]{DM-H_AI_F1_4F_th}}

\caption{Differential cross sections $d\sigma/dT$ for $\chi+\mathrm{H}\rightarrow\chi+p+e^{-}$
via a QED-like $\chi$--$e$ interaction with the $U(1)$ gauge boson
of mass $m_{b}$ and interaction strength $g_{\chi e}\,\alpha$. The
results of the approximation schemes KSV and LPA (not shown in (a)--(d))
are in excellent agreement with the full calculation, and only the
full resutls are shown in (e) and (f). \label{fig:dsigma/dT_DM_F1}}
\end{figure}

Instead of a relativistic muon, consider now the atomic ionization
by some non-relativistic, weakly interacting massive particle, $\chi$,
which could be a dark matter (DM) candidate. Suppose this particle
is of galactic origin with a mean velocity $v_{\chi}\sim220/(3\times10^{5})$,
its kinetic energy $\approx\frac{1}{2}\, m_{\chi}\, v_{\chi}^{2}=270\,(\frac{m_{\chi}}{\mathrm{GeV}})\,\mathrm{eV}$;
therefore, in order to ionize a hydrogen, $m_{\chi}\gtrsim60\,\mathrm{MeV}$.
To make use of the general formalism developed in Sec.~\ref{sec:formalism},
we postulate a QED-like fermionic DM-electron ($\chi e$) interaction
in which the new $U(1)$ gauge boson has mass $m_{b}$ and the interaction
strength $\alpha_{\chi e}\equiv g_{\chi e}\,\alpha$. Fig.~\ref{fig:dsigma/dT_DM_F1}
shows the differential cross sections for $m_{\chi}=100\,\mathrm{MeV}$
and $1\,\mathrm{GeV}$, with either a massless gauge boson $m_{b}=0$,
which corresponds to an infinitely-ranged interaction, or a very massive
one $m_{b}=125\,\mathrm{GeV}$, which leads to a extremely short-ranged
interaction.~%
\footnote{Note that the coupling strength $g_{\chi e}$ is associated with the
choice of $m_{b}$.%
}

Not shown in Fig.~\ref{fig:dsigma/dT_DM_F1}(a)--(d) are the results
of the KSV and LPA, as they are in excellent agreement with the full
calculations in all cases illustrated. Accordingly, the failure of
the EPA scheme is anticipated. On the other hand, although it is expected
that the binding effect should suppress the FEA results, the several
orders of magnitude overestimation by the FEA scheme in the entire
range of energy transfer, evidenced in panels (a)--(d), indicates
the inadequacy of the FEA scheme in such a kinematic regime. Figs.~\ref{fig:dsigma/dT_DM_F1}(e)
and (f) show that the differential cross section becomes ``saturated''
when $m_{\chi}$ becomes much bigger than $1\,\mathrm{GeV}$, which
is about the mass of the hydrogen target. This can be understood by
transforming the laboratory frame, where the hydrogen target is stationary,
to the DM rest frame which coincides the center-of-mass frame for
$m_{\chi}\gg m_p$: the kinematics only depends on $v_{\chi}$
and the reduced mass $\approx m_{p}$. Also by comparing
the case with $m_{b}=0$, Figs.~\ref{fig:dsigma/dT_DM_F1}(a,b,e),
and $m_{b}=125\,\mathrm{GeV}$, Figs.~\ref{fig:dsigma/dT_DM_F1}(b,d,e),
the differential cross sections, apart from some overall scale factors,
show a slower decreasing with energy transfer $T$ as the range of
the $\chi e$ interaction decreases.

\begin{figure}
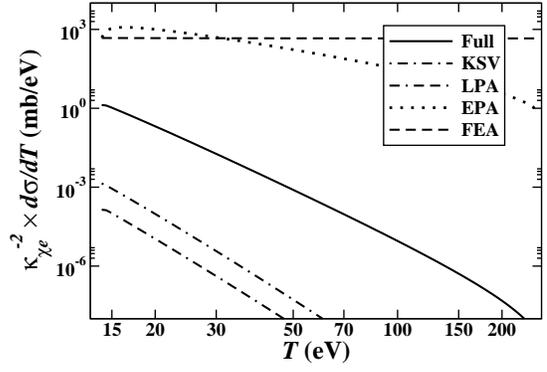
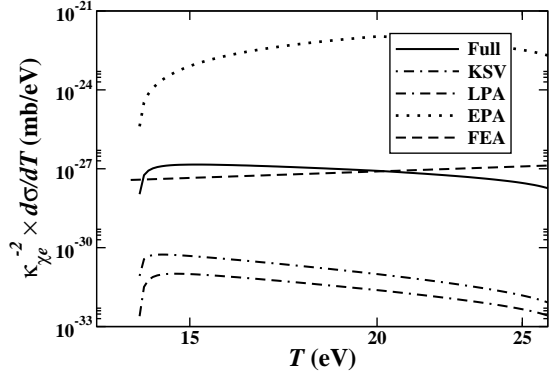
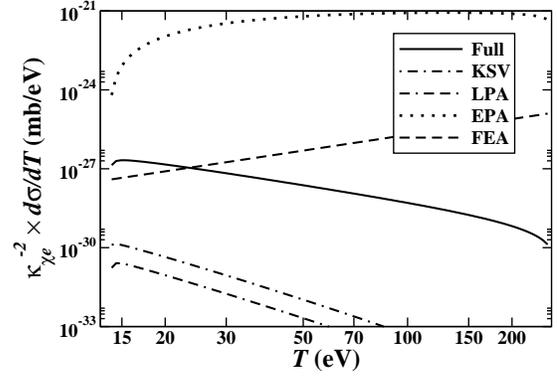
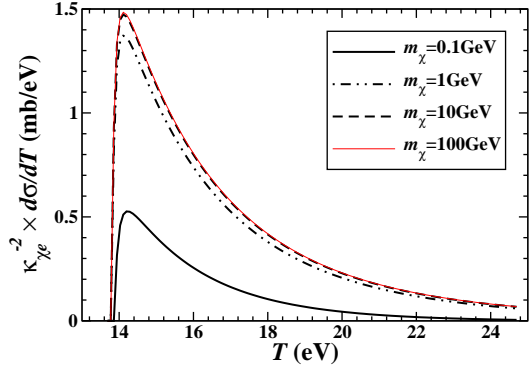
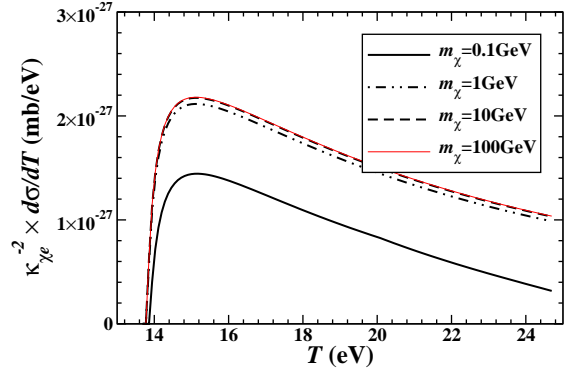

\subfloat[$m_{\chi}=0.1\,\mathrm{GeV}$, $m_b=0$]
{\includegraphics[scale=1]{DM-H_AI_F2_QED_100MeV}}
\hfill
\subfloat[$m_{\chi}=1\,\mathrm{GeV}$, $m_b=0$]
{\includegraphics[scale=1]{DM-H_AI_F2_QED_1GeV}}
\\ \vspace{0.5cm}
\subfloat[$m_{\chi}=0.1\,\mathrm{GeV}$, $m_b=125\,\mathrm{GeV}$]
{\includegraphics[scale=1]{DM-H_AI_F2_4F_100MeV}}
\hfill
\subfloat[$m_{\chi}=1\,\mathrm{GeV}$, $m_b=125\,\mathrm{GeV}$]
{\includegraphics[scale=1]{DM-H_AI_F2_4F_1GeV}}
\\ \vspace{0.5cm}
\subfloat[$m_b=0$, near threshold]
{\includegraphics[scale=1]{DM-H_AI_F2_QED_th}}
\hfill
\subfloat[$m_b=125\,\mathrm{GeV}$, near threshold]
{\includegraphics[scale=1]{DM-H_AI_F2_4F_th}}

\caption{Differential cross sections $d\sigma/dT$ for $\chi+\mathrm{H}\rightarrow\bar{v}+p+e^{-}$
via an anomalous-magnetic-moment-like $\chi$--$e$ interaction with
the $U(1)$ gauge boson of mass $m_{b}$ and coupling strength $\kappa_{\chi e}\,\alpha$.
Only the full results are shown in (e) and (f). \label{fig:dsigma/dT_DM_F2}}
\end{figure}

If, on other hand, the neutral fermionic dark matter has a non-zero
(anomalous) magnetic moment, or its coupling to the $U(1)$ gauge
$ $ boson is via the Dirac bilinear $\bar{\chi}\,\sigma_{\mu\nu}\, q^{\nu}\,\chi/(2\, m_{e})$,
a different constant $\alpha_{\chi e}=\kappa_{\chi e}\,\alpha$ is
assigned to characterize the interaction strength. This anomalous-magnetic-moment-like
interaction yields quite different results, as shown in Fig.~\ref{fig:dsigma/dT_DM_F2},
from the previous case with the same kinematics. The most noticeable
difference seen in Figs.~\ref{fig:dsigma/dT_DM_F2}(a)--(d) is that
the KSV and LPA no longer work, and in fact, largely underestimate.
This implies not only substantial contributions from the transverse
response but also the breakdown of long wavelength approximation,
which has been good for all cases previously discussed. However, the
EPA does not work either: it yields a huge overestimate which implies
the transverse kinematics is not dominated by the photon-like, $Q^{2}\approx0$,
region. Therefore, one encounters a very subtle kinematic regime where
none of the approximation schemes work and requires a full calculation.
While Figs.~\ref{fig:dsigma/dT_DM_F2}(e) and (f) show a similar
cross section saturation for $m_{\chi}\gg m_{p}$ and the
range effect on the differential cross section as previously found,
a comparison of Fig.~\ref{fig:dsigma/dT_DM_F2} and Fig.~\ref{fig:dsigma/dT_DM_F1}
shows that the large energy transfer regime is more suppressed in
the QED-like interaction than the anomalous-magnetic-moment-like interaction.

The general trend observed above for the $\chi e$ cross sections
that the atomic charge operator dominates in the QED-like interaction
while the transverse current operator in the anomalous-magnetic-moment-like
interaction is opposite to what have been concluded for the ionizations
by relativistic muons ($F_{1}$ coupling) and neutrinos ($F_{2}$
coupling). This difference is also partially due to the corresponding
kinematic factors: With $m_{\chi}^{2}\gg|Q^{2}|,q^{2},T^{2}$, they
are
\begin{align}
V_{L}^{(F_{1})}\approx4\, m_{\chi}^{2}\,\frac{Q^{4}}{q^{4}}\,,\qquad & V_{T}^{(F_{1})}\approx2\, m_{\chi}^{2}\,(\frac{|Q^{2}|}{2\, m_{\chi}^{2}}-\frac{T^{2}}{q^{2}})\,;
\end{align}
and
\begin{align}
V_{L}^{(F_{2})}\approx4\, m_{\chi}^{2}\,\frac{Q^{4}}{q^{4}}\,(\frac{\omega_{1}\, T}{m_{\chi}^{2}}\,|Q^{2}|-T^{2})\,,\qquad & V_{T}^{(F_{2})}\approx2\, m_{\chi}^{2}\,|Q^{2}|\,(1+\frac{|Q^{2}|}{q^{2}})\,.
\end{align}
The square of four momentum transfer in DM scattering is
\begin{equation}
Q^{2}\approx-(2-r_{E}-2\,\sqrt{1-r_{E}}\, x)\, m_{\chi}^{2}\, v_{\chi}^{2}\,,
\end{equation}
where $r_{E}$ is the fraction of the DM kinetic energy transfer to
the atom, i.e., $2\, T/(m_{\chi}\, v_{\chi}^{2})$. For most range
of $x$ (which is less restricted unless $m_{\chi}\gg m_{p}$)
and $r_{E}$ (which can never be zero for ionization), one can estimate
$|Q^{2}|\sim m_{\chi}^{2}\, v_{\chi}^{2}$. Because $T^{2}=(r^{2}\, v_{\chi}^{2}/4)\, m_{\chi}^{2}\, v_{\chi}^{2}$,
$q^{2}=T^{2}+|Q^{2}|\sim m_{\chi}^{2}\, v_{\chi}^{2}$. Using these
estimates, the leading orders in $v_{\chi}$ are $O(1)$ for $V_{L}^{(F_{1})}$,
$O(v_{\chi}^{2})$ for $V_{L}^{(F_{1})}$; and $O(v_{\chi}^{4})$
for $V_{L}^{(F_{2})}$, $O(v_{\chi}^{2})$ for $V_{T}^{(F_{2})}$,
respectively. Therefore, in ratio to the charge operator, the transverse
current is suppressed by $O(v_{\chi}^{2})/O(1)$ in the $F_{1}$-type
coupling, while it is enhanced by $O(v_{\chi}^{2})/O(v_{\chi}^{4})$
in the $F_{2}$-type coupling due to the kinematic factors.

\section{Conclusion \label{sec:summary}}

We studied the ionization of hydrogen by scattering of neutrino magnetic
moment, relativistic muon, and weakly-interacting massive particle
with a QED-like interaction. Analytic results were obtained and compared
with several approximation schemes often used in atomic physics. It
is found that for the case of neutrino magnetic moment, the atomic
charge operator dominates the process, and for typical reactor neutrino
energies about tens of $\mathrm{keV}$ to a few $\mathrm{MeV}$, the
atomic binding effect is negligible. For relativistic muon scattering,
on the other hand, the transverse current operator becomes dominant
with increasing incident muon energy. In this case, the equivalent
photon approximation yields a reasonable result, however, for further
improvement, the contribution from the charge operator needs to be
taken into account. Also, due to the special weight by kinematics,
the free electron approximation largely underestimates the result.
The WIMP scattering is the most kinematics-sensitive case, and the 
free electron approximation fails badly. Depending on the coupling
to the dark matter particle, the cross section is dominated by the
charge operator for the $F_{1}$-coupling, and the transverse current
operator for the $F_{2}$-coupling. While the longitudinal photon approximation
works for the former, none of the approximations under study work
for the latter. 

\begin{acknowledgments}
We thank Henry T. Wong for stimulating discussions and comments. The
work is supported in part by the NSC of ROC under grants 99-2112-M-002-010-MY3, 102-2112-M-002-013-MY3 (JWC, CFL, CLW) and 98-2112-M-259-004-MY3,  101-2112-M-259-001 (CPL).
\end{acknowledgments}

\bibliographystyle{apsrev4-1}
\bibliography{draft_v3}

\begin{thebibliography}{23}%
\makeatletter
\providecommand \@ifxundefined [1]{%
 \@ifx{#1\undefined}
}%
\providecommand \@ifnum [1]{%
 \ifnum #1\expandafter \@firstoftwo
 \else \expandafter \@secondoftwo
 \fi
}%
\providecommand \@ifx [1]{%
 \ifx #1\expandafter \@firstoftwo
 \else \expandafter \@secondoftwo
 \fi
}%
\providecommand \natexlab [1]{#1}%
\providecommand \enquote  [1]{``#1''}%
\providecommand \bibnamefont  [1]{#1}%
\providecommand \bibfnamefont [1]{#1}%
\providecommand \citenamefont [1]{#1}%
\providecommand \href@noop [0]{\@secondoftwo}%
\providecommand \href [0]{\begingroup \@sanitize@url \@href}%
\providecommand \@href[1]{\@@startlink{#1}\@@href}%
\providecommand \@@href[1]{\endgroup#1\@@endlink}%
\providecommand \@sanitize@url [0]{\catcode `\\12\catcode `\$12\catcode
  `\&12\catcode `\#12\catcode `\^12\catcode `\_12\catcode `\%12\relax}%
\providecommand \@@startlink[1]{}%
\providecommand \@@endlink[0]{}%
\providecommand \url  [0]{\begingroup\@sanitize@url \@url }%
\providecommand \@url [1]{\endgroup\@href {#1}{\urlprefix }}%
\providecommand \urlprefix  [0]{URL }%
\providecommand \Eprint [0]{\href }%
\providecommand \doibase [0]{http://dx.doi.org/}%
\providecommand \selectlanguage [0]{\@gobble}%
\providecommand \bibinfo  [0]{\@secondoftwo}%
\providecommand \bibfield  [0]{\@secondoftwo}%
\providecommand \translation [1]{[#1]}%
\providecommand \BibitemOpen [0]{}%
\providecommand \bibitemStop [0]{}%
\providecommand \bibitemNoStop [0]{.\EOS\space}%
\providecommand \EOS [0]{\spacefactor3000\relax}%
\providecommand \BibitemShut  [1]{\csname bibitem#1\endcsname}%
\let\auto@bib@innerbib\@empty
\bibitem [{\citenamefont {Beringer}\ \emph {et~al.}(2012)\citenamefont
  {Beringer} \emph {et~al.}}]{Beringer:1900zz}%
  \BibitemOpen
  \bibfield  {author} {\bibinfo {author} {\bibfnamefont {J.}~\bibnamefont
  {Beringer}} \emph {et~al.} (\bibinfo {collaboration} {Particle Data Group}),\
  }\href {\doibase 10.1103/PhysRevD.86.010001} {\bibfield  {journal} {\bibinfo
  {journal} {Phys. Rev. D}\ }\textbf {\bibinfo {volume} {86}},\ \bibinfo
  {pages} {010001} (\bibinfo {year} {2012})}\BibitemShut {NoStop}%
\bibitem [{\citenamefont {Broggini}\ \emph {et~al.}(2012)\citenamefont
  {Broggini}, \citenamefont {Giunti},\ and\ \citenamefont
  {Studenikin}}]{Broggini:2012df}%
  \BibitemOpen
  \bibfield  {author} {\bibinfo {author} {\bibfnamefont {C.}~\bibnamefont
  {Broggini}}, \bibinfo {author} {\bibfnamefont {C.}~\bibnamefont {Giunti}}, \
  and\ \bibinfo {author} {\bibfnamefont {A.}~\bibnamefont {Studenikin}},\
  }\href {\doibase 10.1155/2012/459526} {\bibfield  {journal} {\bibinfo
  {journal} {Adv. High Energy Phys.}\ }\textbf {\bibinfo {volume} {2012}},\
  \bibinfo {pages} {459526} (\bibinfo {year} {2012})}\BibitemShut {NoStop}%
\bibitem [{\citenamefont {Aseev}\ \emph {et~al.}(2011)\citenamefont {Aseev}
  \emph {et~al.}}]{Aseev:2011dq}%
  \BibitemOpen
  \bibfield  {author} {\bibinfo {author} {\bibfnamefont {V.~N.}\ \bibnamefont
  {Aseev}} \emph {et~al.} (\bibinfo {collaboration} {Troitsk Collaboration}),\
  }\href {\doibase 10.1103/PhysRevD.84.112003} {\bibfield  {journal} {\bibinfo
  {journal} {Phys. Rev. D}\ }\textbf {\bibinfo {volume} {84}},\ \bibinfo
  {pages} {112003} (\bibinfo {year} {2011})}\BibitemShut {NoStop}%
\bibitem [{\citenamefont {Beda}\ \emph {et~al.}(2013)\citenamefont {Beda},
  \citenamefont {Brudanin}, \citenamefont {Egorov}, \citenamefont {Medvedev},
  \citenamefont {Pogosov} \emph {et~al.}}]{Beda:2013mta}%
  \BibitemOpen
  \bibfield  {author} {\bibinfo {author} {\bibfnamefont {A.~G.}\ \bibnamefont
  {Beda}}, \bibinfo {author} {\bibfnamefont {V.~B.}\ \bibnamefont {Brudanin}},
  \bibinfo {author} {\bibfnamefont {V.~G.}\ \bibnamefont {Egorov}}, \bibinfo
  {author} {\bibfnamefont {D.~V.}\ \bibnamefont {Medvedev}}, \bibinfo {author}
  {\bibfnamefont {V.~S.}\ \bibnamefont {Pogosov}},  \emph {et~al.},\ }\href
  {\doibase 10.1134/S1547477113020027} {\bibfield  {journal} {\bibinfo
  {journal} {Phys. Part. Nucl. Lett.}\ }\textbf {\bibinfo {volume} {10}},\
  \bibinfo {pages} {139} (\bibinfo {year} {2013})}\BibitemShut {NoStop}%
\bibitem [{\citenamefont {Wong}\ \emph {et~al.}(2007)\citenamefont {Wong} \emph
  {et~al.}}]{Wong:2006nx}%
  \BibitemOpen
  \bibfield  {author} {\bibinfo {author} {\bibfnamefont {H.~T.}\ \bibnamefont
  {Wong}} \emph {et~al.} (\bibinfo {collaboration} {TEXONO}),\ }\href {\doibase
  10.1103/PhysRevD.75.012001} {\bibfield  {journal} {\bibinfo  {journal} {Phys.
  Rev. D}\ }\textbf {\bibinfo {volume} {75}},\ \bibinfo {pages} {012001}
  (\bibinfo {year} {2007})}\BibitemShut {NoStop}%
\bibitem [{\citenamefont {Arpesella}\ \emph {et~al.}(2008)\citenamefont
  {Arpesella} \emph {et~al.}}]{Arpesella:2008mt}%
  \BibitemOpen
  \bibfield  {author} {\bibinfo {author} {\bibfnamefont {C.}~\bibnamefont
  {Arpesella}} \emph {et~al.} (\bibinfo {collaboration} {The Borexino
  Collaboration}),\ }\href {\doibase 10.1103/PhysRevLett.101.091302} {\bibfield
   {journal} {\bibinfo  {journal} {Phys. Rev. Lett.}\ }\textbf {\bibinfo
  {volume} {101}},\ \bibinfo {pages} {091302} (\bibinfo {year}
  {2008})}\BibitemShut {NoStop}%
\bibitem [{\citenamefont {Vogel}\ and\ \citenamefont
  {Engel}(1989)}]{Vogel:1989iv}%
  \BibitemOpen
  \bibfield  {author} {\bibinfo {author} {\bibfnamefont {P.}~\bibnamefont
  {Vogel}}\ and\ \bibinfo {author} {\bibfnamefont {J.}~\bibnamefont {Engel}},\
  }\href {\doibase 10.1103/PhysRevD.39.3378} {\bibfield  {journal} {\bibinfo
  {journal} {Phys. Rev. D}\ }\textbf {\bibinfo {volume} {39}},\ \bibinfo
  {pages} {3378} (\bibinfo {year} {1989})}\BibitemShut {NoStop}%
\bibitem [{\citenamefont {Wong}(2011)}]{Wong:2011zzd}%
  \BibitemOpen
  \bibfield  {author} {\bibinfo {author} {\bibfnamefont {H.~T.}\ \bibnamefont
  {Wong}},\ }\href {\doibase 10.1088/1742-6596/309/1/012024} {\bibfield
  {journal} {\bibinfo  {journal} {J. Phys. Conf. Ser.}\ }\textbf {\bibinfo
  {volume} {309}},\ \bibinfo {pages} {012024} (\bibinfo {year}
  {2011})}\BibitemShut {NoStop}%
\bibitem [{\citenamefont {Yue}\ and\ \citenamefont {Wong}(2013)}]{Yue:2013jja}%
  \BibitemOpen
  \bibfield  {author} {\bibinfo {author} {\bibfnamefont {Q.}~\bibnamefont
  {Yue}}\ and\ \bibinfo {author} {\bibfnamefont {H.~T.}\ \bibnamefont {Wong}},\
  }\href {\doibase 10.1142/S0217732313400075} {\bibfield  {journal} {\bibinfo
  {journal} {Mod. Phys. Lett. A}\ }\textbf {\bibinfo {volume} {28}},\ \bibinfo
  {pages} {1340007} (\bibinfo {year} {2013})}\BibitemShut {NoStop}%
\bibitem [{\citenamefont {Wong}\ \emph {et~al.}(2010)\citenamefont {Wong},
  \citenamefont {Li},\ and\ \citenamefont {Lin}}]{Wong:2010pb}%
  \BibitemOpen
  \bibfield  {author} {\bibinfo {author} {\bibfnamefont {H.~T.}\ \bibnamefont
  {Wong}}, \bibinfo {author} {\bibfnamefont {H.-B.}\ \bibnamefont {Li}}, \ and\
  \bibinfo {author} {\bibfnamefont {S.-T.}\ \bibnamefont {Lin}},\ }\href
  {\doibase 10.1103/PhysRevLett.105.061801} {\bibfield  {journal} {\bibinfo
  {journal} {Phys. Rev. Lett.}\ }\textbf {\bibinfo {volume} {105}},\ \bibinfo
  {pages} {061801} (\bibinfo {year} {2010})},\ \bibinfo {note} {erratum:
  arXiv:1001.2074v3}\BibitemShut {NoStop}%
\bibitem [{\citenamefont {Voloshin}(2010)}]{Voloshin:2010vm}%
  \BibitemOpen
  \bibfield  {author} {\bibinfo {author} {\bibfnamefont {M.~B.}\ \bibnamefont
  {Voloshin}},\ }\href {\doibase 10.1103/PhysRevLett.105.201801} {\bibfield
  {journal} {\bibinfo  {journal} {Phys. Rev. Lett.}\ }\textbf {\bibinfo
  {volume} {105}},\ \bibinfo {pages} {201801} (\bibinfo {year} {2010})},\
  \bibinfo {note} {erratum: {\it ibid}. {\bf 106}, 059901 (2011)}\BibitemShut
  {NoStop}%
\bibitem [{\citenamefont {Kouzakov}\ and\ \citenamefont
  {Studenikin}(2011)}]{Kouzakov:2010tx}%
  \BibitemOpen
  \bibfield  {author} {\bibinfo {author} {\bibfnamefont {K.~A.}\ \bibnamefont
  {Kouzakov}}\ and\ \bibinfo {author} {\bibfnamefont {A.~I.}\ \bibnamefont
  {Studenikin}},\ }\href {\doibase 10.1016/j.physletb.2010.12.043} {\bibfield
  {journal} {\bibinfo  {journal} {Phys. Lett. B}\ }\textbf {\bibinfo {volume}
  {696}},\ \bibinfo {pages} {252} (\bibinfo {year} {2011})}\BibitemShut
  {NoStop}%
\bibitem [{\citenamefont {Kouzakov}\ \emph {et~al.}(2011)\citenamefont
  {Kouzakov}, \citenamefont {Studenikin},\ and\ \citenamefont
  {Voloshin}}]{Kouzakov:2011vx}%
  \BibitemOpen
  \bibfield  {author} {\bibinfo {author} {\bibfnamefont {K.~A.}\ \bibnamefont
  {Kouzakov}}, \bibinfo {author} {\bibfnamefont {A.~I.}\ \bibnamefont
  {Studenikin}}, \ and\ \bibinfo {author} {\bibfnamefont {M.~B.}\ \bibnamefont
  {Voloshin}},\ }\href {\doibase 10.1103/PhysRevD.83.113001} {\bibfield
  {journal} {\bibinfo  {journal} {Phys. Rev. D}\ }\textbf {\bibinfo {volume}
  {83}},\ \bibinfo {pages} {113001} (\bibinfo {year} {2011})}\BibitemShut
  {NoStop}%
\bibitem [{\citenamefont {Essig}\ \emph
  {et~al.}(2012{\natexlab{a}})\citenamefont {Essig}, \citenamefont {Mardon},\
  and\ \citenamefont {Volansky}}]{Essig:2011nj}%
  \BibitemOpen
  \bibfield  {author} {\bibinfo {author} {\bibfnamefont {R.}~\bibnamefont
  {Essig}}, \bibinfo {author} {\bibfnamefont {J.}~\bibnamefont {Mardon}}, \
  and\ \bibinfo {author} {\bibfnamefont {T.}~\bibnamefont {Volansky}},\ }\href
  {\doibase 10.1103/PhysRevD.85.076007} {\bibfield  {journal} {\bibinfo
  {journal} {Phys. Rev. D}\ }\textbf {\bibinfo {volume} {85}},\ \bibinfo
  {pages} {076007} (\bibinfo {year} {2012}{\natexlab{a}})}\BibitemShut
  {NoStop}%
\bibitem [{\citenamefont {Essig}\ \emph
  {et~al.}(2012{\natexlab{b}})\citenamefont {Essig}, \citenamefont
  {Manalaysay}, \citenamefont {Mardon}, \citenamefont {Sorensen},\ and\
  \citenamefont {Volansky}}]{Essig:2012yx}%
  \BibitemOpen
  \bibfield  {author} {\bibinfo {author} {\bibfnamefont {R.}~\bibnamefont
  {Essig}}, \bibinfo {author} {\bibfnamefont {A.}~\bibnamefont {Manalaysay}},
  \bibinfo {author} {\bibfnamefont {J.}~\bibnamefont {Mardon}}, \bibinfo
  {author} {\bibfnamefont {P.}~\bibnamefont {Sorensen}}, \ and\ \bibinfo
  {author} {\bibfnamefont {T.}~\bibnamefont {Volansky}},\ }\href {\doibase
  10.1103/PhysRevLett.109.021301} {\bibfield  {journal} {\bibinfo  {journal}
  {Phys. Rev. Lett.}\ }\textbf {\bibinfo {volume} {109}},\ \bibinfo {pages}
  {021301} (\bibinfo {year} {2012}{\natexlab{b}})}\BibitemShut {NoStop}%
\bibitem [{\citenamefont {{Nordsieck}}(1954)}]{Nordsieck:1953aa}%
  \BibitemOpen
  \bibfield  {author} {\bibinfo {author} {\bibfnamefont {A.}~\bibnamefont
  {{Nordsieck}}},\ }\href {\doibase 10.1103/PhysRev.93.785} {\bibfield
  {journal} {\bibinfo  {journal} {Phys. Rev.}\ }\textbf {\bibinfo {volume}
  {93}},\ \bibinfo {pages} {785} (\bibinfo {year} {1954})}\BibitemShut
  {NoStop}%
\bibitem [{\citenamefont {{Holt}}(1969)}]{Holt:1969ar}%
  \BibitemOpen
  \bibfield  {author} {\bibinfo {author} {\bibfnamefont {A.~R.}\ \bibnamefont
  {{Holt}}},\ }\href {\doibase 10.1088/0022-3700/2/11/311} {\bibfield
  {journal} {\bibinfo  {journal} {J. Phys. B}\ }\textbf {\bibinfo {volume}
  {2}},\ \bibinfo {pages} {1209} (\bibinfo {year} {1969})}\BibitemShut
  {NoStop}%
\bibitem [{\citenamefont {{Belki\'c}}(1981)}]{Belkic:1981dz}%
  \BibitemOpen
  \bibfield  {author} {\bibinfo {author} {\bibfnamefont {D.}~\bibnamefont
  {{Belki\'c}}},\ }\href {\doibase 10.1088/0022-3700/14/12/005} {\bibfield
  {journal} {\bibinfo  {journal} {J. Phys. B}\ }\textbf {\bibinfo {volume}
  {14}},\ \bibinfo {pages} {1907} (\bibinfo {year} {1981})}\BibitemShut
  {NoStop}%
\bibitem [{\citenamefont {{Gravielle}}\ and\ \citenamefont
  {{Miraglia}}(1992)}]{Gravielle:1991mi}%
  \BibitemOpen
  \bibfield  {author} {\bibinfo {author} {\bibfnamefont {M.~S.}\ \bibnamefont
  {{Gravielle}}}\ and\ \bibinfo {author} {\bibfnamefont {J.~E.}\ \bibnamefont
  {{Miraglia}}},\ }\href {\doibase 10.1016/0010-4655(92)90127-K} {\bibfield
  {journal} {\bibinfo  {journal} {Comp. Phys. Comm.}\ }\textbf {\bibinfo
  {volume} {69}},\ \bibinfo {pages} {53} (\bibinfo {year} {1992})}\BibitemShut
  {NoStop}%
\bibitem [{\citenamefont {von Weizsacker}(1934)}]{Weizsacker:1934sx}%
  \BibitemOpen
  \bibfield  {author} {\bibinfo {author} {\bibfnamefont {C.~F.}\ \bibnamefont
  {von Weizsacker}},\ }\href {\doibase 10.1007/BF01333110} {\bibfield
  {journal} {\bibinfo  {journal} {Z. Phys.}\ }\textbf {\bibinfo {volume}
  {88}},\ \bibinfo {pages} {612} (\bibinfo {year} {1934})}\BibitemShut
  {NoStop}%
\bibitem [{\citenamefont {Williams}(1934)}]{Williams:1934ad}%
  \BibitemOpen
  \bibfield  {author} {\bibinfo {author} {\bibfnamefont {E.~J.}\ \bibnamefont
  {Williams}},\ }\href {\doibase 10.1103/PhysRev.45.729} {\bibfield  {journal}
  {\bibinfo  {journal} {Phys. Rev.}\ }\textbf {\bibinfo {volume} {45}},\
  \bibinfo {pages} {729} (\bibinfo {year} {1934})}\BibitemShut {NoStop}%
\bibitem [{\citenamefont {Siegert}(1937)}]{Siegert:1937yt}%
  \BibitemOpen
  \bibfield  {author} {\bibinfo {author} {\bibfnamefont {A.~J.~F.}\
  \bibnamefont {Siegert}},\ }\href {\doibase 10.1103/PhysRev.52.787} {\bibfield
   {journal} {\bibinfo  {journal} {Phys. Rev.}\ }\textbf {\bibinfo {volume}
  {52}},\ \bibinfo {pages} {787} (\bibinfo {year} {1937})}\BibitemShut
  {NoStop}%
\bibitem [{\citenamefont {Landau}\ and\ \citenamefont
  {Lifshitz}(1981)}]{Landau:NQM}%
  \BibitemOpen
  \bibfield  {author} {\bibinfo {author} {\bibfnamefont {L.~D.}\ \bibnamefont
  {Landau}}\ and\ \bibinfo {author} {\bibfnamefont {L.~M.}\ \bibnamefont
  {Lifshitz}},\ }\href@noop {} {\emph {\bibinfo {title} {Quantum Mechanics:
  Non-Relativistic Theory}}},\ \bibinfo {edition} {3rd}\ ed.\ (\bibinfo
  {publisher} {Butterworth-Heinemann},\ \bibinfo {year} {1981})\BibitemShut
  {NoStop}%
\end{thebibliography}%

\end{document}